# Deep Learning in Proteomics Informatics: Applications, Challenges, and Future Directions

**Yindan Luo**[+]

School of Computer and Imformation Engineering

Xiamen University of Technology

Xiamen, China

1015959737@qq.com

**Jiaxin Cai**[+*]

School of Mathematics and Statistics

Xiamen University of Technology

Xiamen, China

caijiaxin@xmut.edu.cn

[*] Corresponding author: Jiaxin Cai.

[+] Yindan Luo and Jiaxin Cai are co-first authors.

**Abstract:** Deep learning is an advanced technology that relies on large-scale data and complex models for feature extraction and pattern recognition. It has been widely applied across various fields, including computer vision, natural language processing, and speech recognition. In recent years, deep learning has demonstrated significant potential in the realm of proteomics informatics, particularly in deciphering complex biological information. The introduction of this technology not only accelerates the processing speed of protein data but also enhances the accuracy of predictions regarding protein structure and function. This provides robust support for both fundamental biology research and applied biotechnological studies. Currently, deep learning is primarily focused on applications such as protein sequence analysis, three-dimensional structure prediction, functional annotation, and the construction of protein interaction networks. These applications offer numerous advantages to proteomic research. Despite its growing prevalence in this field, deep learning faces several challenges including data scarcity, insufficient model interpretability, and computational complexity; these factors hinder its further advancement within proteomics. This paper comprehensively reviews the applications of deep learning in proteomics along with the challenges it encounters. The aim is to provide a systematic theoretical discussion and practical basis for research in this domain to facilitate ongoing development and innovation of deep learning technologies within proteomics.

## I. INTRODUCTION

Deep learning, a prominent subset of machine learning, processes and analyzes data by emulating the neural network structures found in the human brain, utilizing multi-layered network architectures [1]. The basic architecture of deep learning models consists of an input layer, one or more hidden layers, and an output layer [2]. Although deep learning has garnered substantial attention in recent years, its conceptual origins can be traced back to 1943, when neuroscientist Warren McCulloch and mathematician Walter Pitts introduced a model for artificial neural networks [3]. Deep learning did not experience significant advancements until the early 21st century, primarily due to enhancements in computational power and the proliferation of large-scale data resources. Since then, this technology has been widely applied across diverse domains, including image recognition, natural language processing, autonomous driving, and bioinformatics [4]. These advancements have established a robust foundation for accelerated progress in the application of deep learning within the field of proteomics.Deep learning models predominantly comprise several types, such as deep neural networks [5-6], convolutional neural networks [7], and recurrent neural networks [8-9]. These models are designed to extract and learn features from data via a multi-layered network architecture. Owing to the remarkable flexibility and adaptability of deep learning, it has been extensively applied across various domains, especially in tasks related to the automatic recognition, classification, and prediction of complex data patterns, where it exhibits considerable advantages.

In the domain of proteomics [10], the burgeoning volume of proteomic data presents a substantial challenge for scientific research in terms of efficient extraction, analysis, and interpretation. Conventional data analysis methodologies are not only labor-intensive but also susceptible to subjective biases. Moreover, they frequently fall short in elucidating the intricate patterns and associations inherent within the data [11]. The advent of deep learning technologies, characterized by their robust data processing capabilities and automated feature extraction methods, has opened new avenues for overcoming these bottlenecks, thereby substantially improving the efficiency of data processing and analysis. Deep learning is predominantly utilized in diverse areas, including protein sequence analysis, three-dimensional structure prediction, functional annotation, and interaction studies, and its applications further extend to genomics [12] and personalized medicine. For example, in the realm of protein structure prediction, deep learning methodologies such as AlphaFold [13] have exhibited significant superiority over conventional techniques like homology modeling and conformational search methods by accurately forecasting protein folding structures. This advancement offers vital insights for drug design and pathological research.

Despite the promising potential of deep learning in the analysis of biomedical data, it concurrently encounters a range of challenges and limitations. Specifically, in the context of multimodal data processing [14], various types of biological information—such as genomics, transcriptomics, and proteomics—are often characterized by diverse forms and structures. Analyzing a single data type may be insufficient to fully capture the complexity of biological systems. Consequently, there is an urgent need to develop more effective strategies for data integration. This article aims to offer a comprehensive review of recent advancements in proteomics informatics through the lens of deep learning, with a particular emphasis on its applications in this domain. Such advancements not only introduce novel perspectives and tools for biomedical research but also propel the progress of multimodal deep learning technologies, thereby creating new avenues for future medical research and clinical applications. While contemporary deep learning algorithms have yet to achieve an optimal state, they are instrumental in the progression of bioinformatics and medical research. The structure of this paper is organized as follows: Section II provides an overview of the developmental background of deep learning; Section III examines the current status and potential applications of deep learning in proteomics informatics; Section IV analyzes the challenges, limitations, and risks associated with deep learning; and Section V explores future directions for deep learning in proteomics informatics; and Section VI provides a summary of the entire paper. This paper focuses on recent deep learning advancements in proteomics amid rapid AI progress, serving as a reference and inspiration for researchers, rather than an exhaustive survey.

## II. BACKGROUND OF DEEP LEARNING

In recent years, the exponential increase in data volume has underscored the advantages of deep neural networks, thereby significantly accelerating the advancement of deep learning technologies. These networks are endowed with the ability to automatically extract features, a trait that is of considerable value in biological applications [15]. Numerous protein prediction methods using deep learning have been created. This section analyzes six common deep neural network architectures used in deep learning, with Figure I comparing and summarizing various models.

### *A. Deep Neural Network*

In 2006, Hinton and colleagues [16] published a seminal paper in the journal Science that introduced deep neural networks, heralding the advent of deep learning. Deep neural networks (DNNs) are advanced feedforward neural networks based on multilayer perceptrons, inspired by the human brain's neural architecture. Unlike shallow networks, DNNs consist of an input layer, multiple hidden layers, and an output layer, allowing them to process data through nonlinear transformations to learn complex features and extract valuable information. This method reduces reliance on manual feature engineering by adjusting weights to produce outputs. During DNN training, the backpropagation algorithm is used to update network weights layer by layer, optimizing the model by minimizing the difference between the loss function and actual outputs. DNNs excel in fields like computer vision, natural language processing, and medical image analysis. In proteomics, they effectively extract complex features from amino acid sequences using multi-layer networks, enabling accurate predictions of protein interactions.

### *B. Convolutional Neural Network*

The introduction of the LeNet-5 convolutional neural network [17] represents the formal beginning of CNN architectures. Since the groundbreaking development of AlexNet [18] in 2012, CNNs are a key part of deep learning, designed to handle grid-like data, especially images. They are composed of convolutional, pooling, and fully connected layers. The convolutional layer is tasked with extracting local features from the data [19] and conducting feature extraction, while the pooling layer downsamples feature maps, preserving essential information. Unlike DNNs, CNNs reduce fully connected layers to lower parameter count and computational complexity. They offer three key benefits: local connectivity, weight sharing, and pooling. Local connectivity helps identify feature similarities, and weight sharing cuts down model parameters, further reducing complexity. Concurrently, pooling operations reduce feature space dimensionality. CNNs are popular in computer vision for their robust representation learning and resistance to transformations such as scaling, translation, and rotation. In biomedicine, they excel in image registration and recognition. Early deep learning methods for protein function prediction also utilized CNN architectures.

### *C. Recurrent Neural Network*

Recurrent Neural Networks (RNNs) are deep learning models designed for sequential data analysis. They consist of an input layer, one or more hidden layers, and an output layer. Unlike DNNs and CNNs, RNNs have interconnected inputs and outputs, enabling them to use previous context effectively when processing current data. RNNs can efficiently capture dynamic features and dependencies in time series data, showcasing short-term memory retention. Typically, RNNs are trained using backpropagation algorithms, particularly the Backpropagation Through Time (BPTT) technique [20], to update weight matrices. Nonetheless, conventional RNNs frequently face challenges such as vanishing or exploding gradients when learning from extended sequences, which can result in reduced efficacy in capturing long-term dependencies. To overcome these limitations, researchers have introduced several variants of RNNs, notably Long Short-Term Memory (LSTM) networks [21] and Gated Recurrent Units (GRUs) [22]. These variants use gating mechanisms to better manage information flow and improve handling of long-range dependencies. As a result, RNNs and their advanced versions are widely used in fields like natural language processing, computer vision, and computational biology, particularly for identifying cases and predicting protein subcellular localization.

### *D. Graph Neural Network*

The inception of Graph Neural Networks (GNNs) can be traced to 2005, when Gori et al. [23] initially introduced the concept. GNNs are neural networks designed for graph data, effectively capturing node relationships and attributes. They are used in fields like natural language processing, image processing, and drug development. Presently, prominent algorithms within the GNN domain include Graph Convolutional Networks (GCNs) [24] and Graph Attention Networks (GATs) [25]. Traditional neural networks struggle with irregular non-Euclidean data, while GCNs, a type of convolutional neural network, excel in processing graph-structured data by aggregating features from neighboring nodes. This capability makes GCNs popular in fields like biological data analysis and knowledge graph construction. A key limitation of GCNs is their uniform weighting of neighboring nodes, which hinders capturing complex node relationships. To overcome this, GATs were developed, incorporating an attention mechanism to dynamically adjust node influence. This spatial approach enables neural networks to focus on important nodes and edges, improving training efficiency and model interpretability. Analyzing graph data helps us better understand complex relationships in unstructured data, which is crucial for studying residue interactions in protein structures.

### *E. Generative Adversarial Network*

Deep learning research has traditionally been closely linked with discriminative models. Nonetheless, in 2014, Ian Goodfellow and his colleagues [26] made a seminal contribution by proposing an unsupervised learning approach termed Generative Adversarial Networks (GANs). This framework aims to generate data resembling real data through competitive and collaborative training. The GAN architecture consists of two main parts: the generator, which creates realistic data from random noise, and the discriminator, which determines if the data is real or generated. These networks undergo iterative training, constantly improving their parameters. This adversarial process helps the generator learn progressively, producing high-quality outputs that test the discriminator's ability to assess authenticity. Unlike traditional deep learning models, GANs excel at capturing real-world data patterns, allowing them to generate highly realistic synthetic data. However, the imbalance between the generator and discriminator in GANs can lead to training instability and gradient vanishing. Nonetheless, GANs excel in image segmentation, style transfer, and data augmentation, and hold potential for predicting protein functions, particularly in analyzing sequences of proteins with unknown functions.

### *F. Transformer*

The attention mechanism, introduced by Ashish Vaswani and colleagues in 2017 [27], underpins the Transformer deep learning model. Inspired by human vision research, this mechanism efficiently allocates resources by assigning different weights to important information, improving data exchange and transmission. Self-attention, a specialized variant of this attention mechanism [28], is an essential component of the Transformer architecture. Transformers can effectively use contextual information by adaptively focusing on different positions in sequences, unlike traditional RNNs. They excel at capturing long-range dependencies, reducing gradient vanishing issues. A Transformer's core structure includes an encoder and a decoder. The encoder is responsible for extracting features from input sequences and transforming them into continuous context vectors, whereas the decoder utilizes these vectors to generate output sequences. Additionally, Transformers use positional encoding to maintain and utilize positional information, improving computational efficiency. Currently, Transformer models are widely used in fields like natural language processing, computer vision, and protein-protein interaction analysis. The previous text provides a comprehensive overview of six distinct types of deep neural networks, highlighting their historical development and practical applications. Each network is characterized by a unique architecture, making it suitable for specific use cases. Collectively, these networks have significantly contributed to the continuous advancement of deep learning technologies and have played a pivotal role in advancing the field of proteomics.

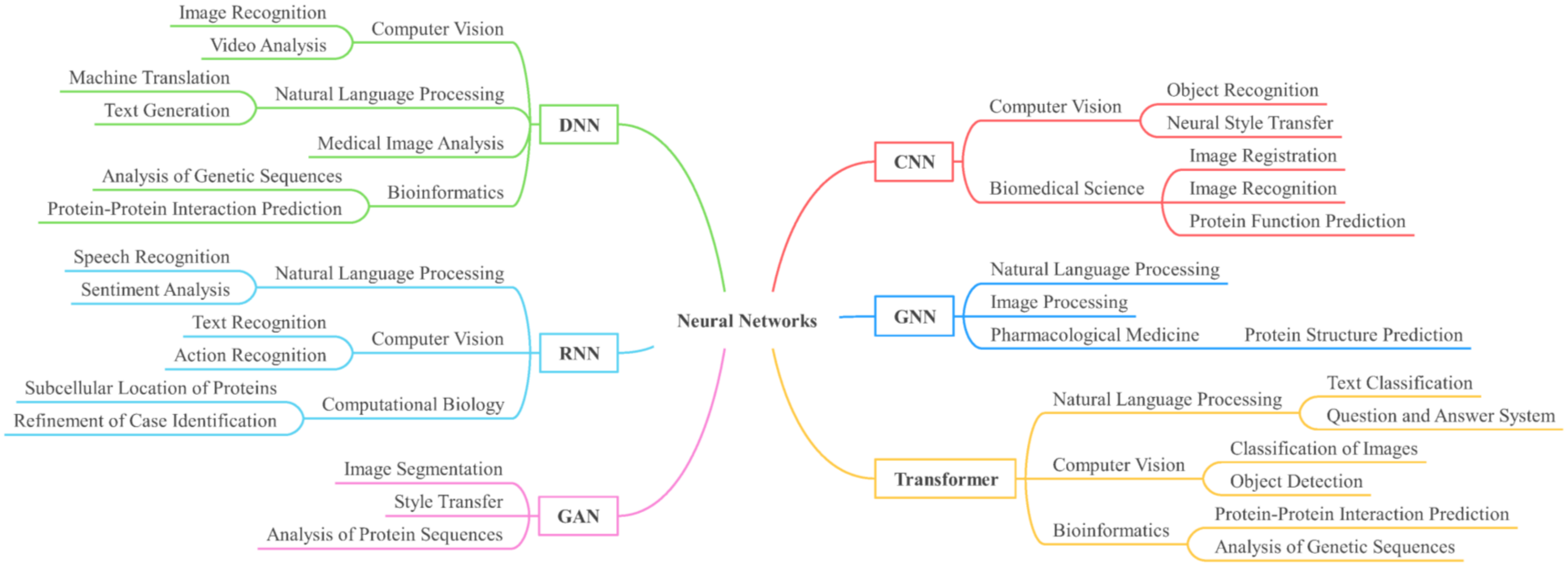

**Fig. I:** The various types of neural networks, encompassing Deep Neural Networks (DNN), Convolutional Neural Networks (CNN), Recurrent Neural Networks (RNN), Graph Neural Networks (GNN), Generative Adversarial Networks (GAN), and Transformers, are extensively applied across multiple domains, including computer vision, natural language processing, and biomedicine.

## III. APPLICATIONS OF DEEP LEARNING IN PROTEOMICS INFORMATICS

### 3.1 Protein Sequence Analysis

Protein sequence analysis constitutes a central research domain within proteomics, encompassing diverse facets such as the prediction, design, and classification of protein sequences (as depicted in Figure II). This field provides an essential foundation for comprehending biological systems. Such analyses facilitate a more profound exploration of the functions and mechanisms of proteins within living organisms. Protein sequence prediction entails the inference of the amino acid sequence from given genetic data. This process is essential for understanding protein structure and function, and it serves as a theoretical basis for drug design and the investigation of disease mechanisms. Recent research has achieved substantial progress in this domain. For example, the work of Repecka et al. [29] introduced ProteinGAN, an innovative generative adversarial network that integrates self-attention mechanisms. This model is capable of discerning evolutionary relationships within protein sequences and generating a wide variety of novel variants exhibiting natural properties. Additionally, the ProGen [30] model represents a significant development, employing approximately 280 million annotated protein sequences for unsupervised learning, which enables the generation of evolutionarily diverse sequences. While these models improve protein sequence prediction, they have limitations. The generated sequences, though diverse, may lack the complex nuances of real biological systems, potentially reducing their accuracy in predicting functionality due to missing context-specific information. Furthermore, the high computational demands for training large-scale models can restrict their accessibility and practical use in the research community.

As AI technology progresses, self-supervised deep language models have excelled in natural language processing. Yet, applying these models to protein sequences fails to fully utilize proteins' unique properties. Recent efforts aim to bridge this gap. For instance, ProteinBERT, as developed by Nadav Brandes et al. [31], represents a deep learning model specifically tailored for protein analysis. It exhibits the ability to adapt to diverse sequence lengths and efficiently handle exceedingly long protein sequences. This capability underscores its potential to address the inherent complexities of protein data, while also demonstrating promising performance across various benchmarks. Another notable contribution is ProteinMPNN, developed by J. Dauparas et al. [32]. This model employs message-passing neural networks [33] to generate protein sequences with innovative structures and functions. The method achieves a sequence recovery rate of 52.4%, significantly surpassing Rosetta's [34] rate of 32.9%, thereby underscoring its superior efficacy in sequence design. Furthermore, GeoSeqBuilder, as developed by Jiale Liu et al. [35], constitutes an innovative deep learning framework that synergistically combines protein sequence generation with side-chain conformation prediction. This integration allows for complex all-atom structural designs, achieving a 51.6% native residue recovery rate for side-chain prediction, with average pLDDT and TM-score values of 78.42 and 0.75 across ten new protein structure datasets. These results match ProteinMPNN's performance and surpass other methods, though further research is necessary. These models show potential for generating protein sequences with desired traits, but their biological relevance and functionality require further validation. There may be discrepancies between predicted and actual protein behaviors, and understanding these deep learning models is challenging. To enhance model performance and scientific insight, it's essential to understand how these models make decisions and what influences specific sequences.

With the advancement of research on protein sequences, the constraints of conventional biological experimental techniques in handling large-scale protein sequence classification have become more apparent. Gu Xingyue [36] proposed a novel neural network methodology utilizing 188-dimensional feature vectors to predict vesicular transport protein sequences. This approach demonstrated substantial progress in the field of protein sequence prediction, achieving precision, accuracy, and recall rates of 0.29, 0.71, and 0.86, respectively, on an independent dataset. Farzana Tasnim et al. [37] employed natural language processing and encoding strategies for the classification of protein sequences. Their results demonstrate that support vector machines [38], when implemented with count vectorization techniques, can attain accuracy, precision, recall, and F1-Score values of 0.92. Furthermore, the robust performance of convolutional neural networks utilizing various encoding methods underscores the potential of these techniques in the analysis of protein sequences. Umesh Kumar Lilhore et al. [39] introduced the ProtICNN-BiLSTM model, which combines an advanced convolutional neural network [40] with attention mechanismsand bidirectional long short-term memory networks [41]. Utilizing Bayesian optimization, this model demonstrated remarkable performance metrics: a specificity of 94.65%, an accuracy of 96.57%, a sensitivity of 95.67%, and a Matthews correlation coefficient of 96.85%. These outcomes are beneficial for the inference of novel protein functions. Despite advancements, limitations remain. Gu Xingyue's method has a low precision rate of 0.29, limiting its use where high precision is crucial. Other methods, though accurate, rely on complex encoding and optimization, complicating implementation and requiring significant computational resources and data preprocessing. The models' ability to generalize across diverse protein sequences and datasets needs more study, as there's a risk of overfitting with varied protein families.

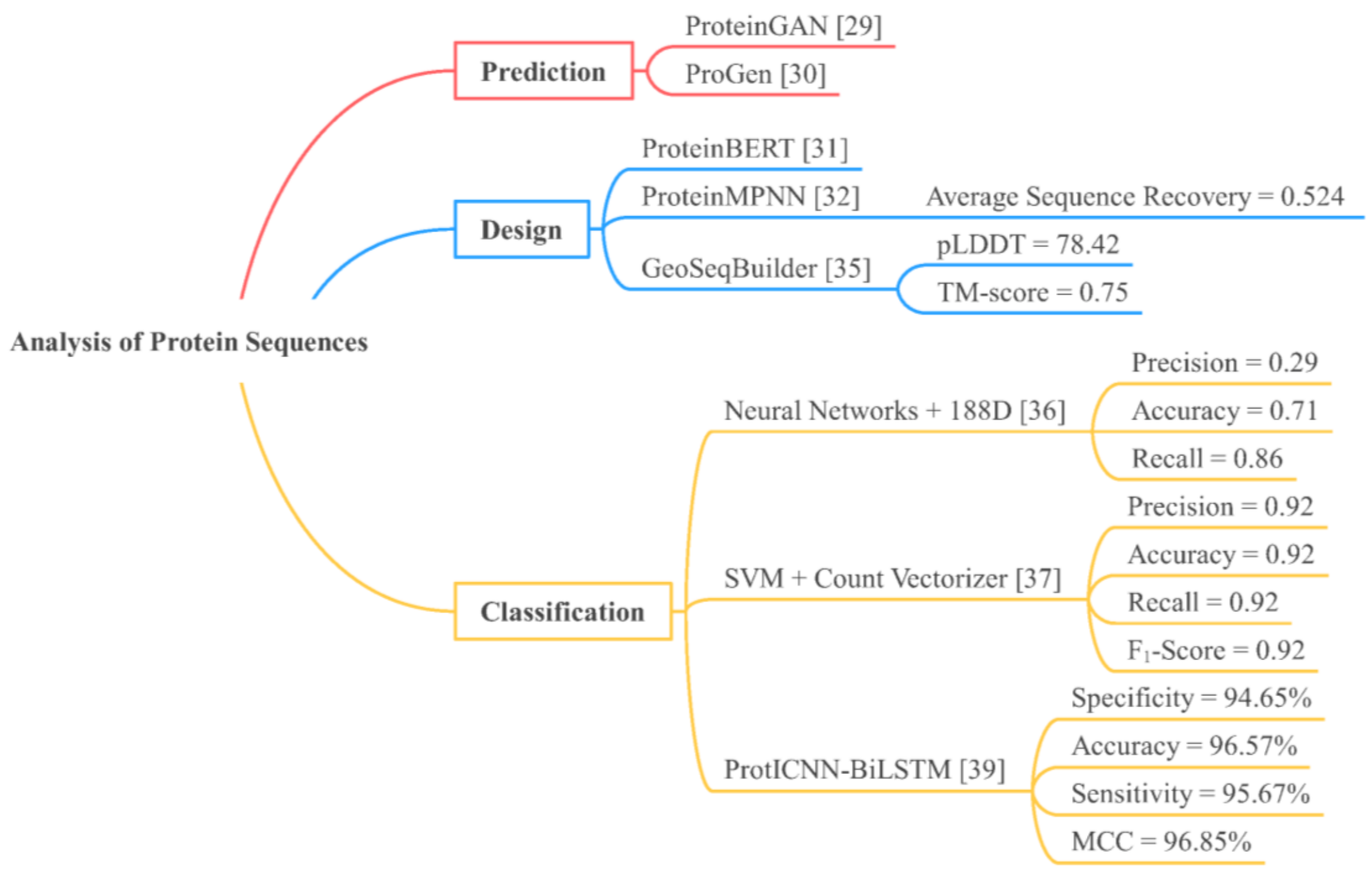

**Fig. II:** The analysis of protein sequences can be divided into three principal components: protein sequence prediction, protein sequence design, and protein sequence classification. Each component employs specific methodologies and tools, each with its own set of performance metrics.

### 3.2 Protein Structure Prediction

Protein structure prediction constitutes a pivotal domain of investigation within proteomics, with its efficacy commonly assessed through the Critical Assessment of Protein Structure Prediction (CASP) [42] and the Critical Assessment of Methods of Protein Structure Prediction (CAMPE) [43]. Detailed information is provided in Table I. Most contemporary deep learning models employ evolutionary data derived from multiple sequence alignments (MSA) [44] to enhance the precision of protein structure prediction. For instance, AlphaFold, created by Andrew W. Senior et al. [45], uses neural networks to predict inter-residue distances and angles, then optimizes a potential energy function with the L-BFGS algorithm. It excels in accuracy even with limited homologous data, achieving a notable TM-score of 0.7 in the CASP13 free modeling evaluation, marking a major advancement in protein structure prediction. Another significant example is trRosetta, a deep residual network introduced by Jianyi Yang et al. [46]. trRosetta employs the Rosetta energy function to predict residue orientations and distances, facilitating rapid and accurate structural modeling. It achieved TM-scores of 0.625 and 0.621 in CASP13 and CAMEO, surpassing many other methods. Additionally, trRosetta excels in de novo design and accurately predicts protein structures without co-evolutionary signals. Recent advancements in deep learning have significantly revolutionized protein structure prediction by improving both the speed and accuracy of the process, thereby facilitating a deeper understanding of protein functions and the design of novel proteins. However, these models face limitations, such as reduced accuracy for complex proteins and reliance on evolutionary data, which can be problematic when such data is unavailable or unreliable.

Co-evolution of amino acids is crucial for estimating inter-residue distances, key for accurate protein structure prediction. While traditional methods use indirect approaches, recent deep learning advancements enable direct distance estimation from multiple sequence alignments. A significant advancement is CopulaNet, a deep neural network by Fusong Ju et.al. [47], designed for accurate residue distance predictions from sequence alignments, aiding protein structure prediction. ProFOLD, incorporating CopulaNet, outperformed AlphaFold in accuracy and TM-score on CASP13 targets, highlighting the potential of new neural network models. The MSA Transformer, created by Roshan Rao et.al. [48], integrates multiple sequence alignment with deep learning attention mechanisms to improve protein structure prediction. It outperformed trRosetta in CASP13 and CAMEO benchmarks, showcasing the power of advanced machine learning in this field. Recent research demonstrates that novel neural network architectures and techniques are significantly improving predictive accuracy. Specifically, CopulaNet advances residue distance predictions, while the MSA Transformer effectively integrates multiple sequence alignment with deep learning methodologies, offering valuable insights for future scholarly investigations.

Through continuous technological innovations and performance optimizations, the field of predictive modeling has witnessed substantial advancements, leading to enhanced accuracy and efficiency. On July 16, 2021, the unveiling of two pioneering models, AlphaFold2 and RoseTTAFold, represented a significant breakthrough in protein structure prediction. AlphaFold2, developed by DeepMind, successfully integrated principles from physics and bioinformatics to attain experimental-level accuracy, outperforming other methodologies in the CASP14 competition. This notable accomplishment was acknowledged through the conferment of the 2024 Nobel Prize in Chemistry and is expected to catalyze further progress in drug design. The RoseTTAFold model, a three-track neural network developed by Minkyung Baek et.al. [49], effectively tackled key challenges related to crystallography and cryo-electron microscopy by swiftly producing models of protein-protein complexes from sequence data. Despite ranking slightly below AlphaFold2 in the CASP14 competition, it nonetheless constituted a substantial advancement in the field of protein structure prediction. Recent advancements in deep learning models, such as DMPfold2 [50], Fold [51], OpenFold [52], and PaddlePaddle [53], have significantly improved the efficiency and accessibility of protein structure prediction in comparison to AlphaFold2 and RoseTTAFold. These contemporary models demonstrate enhanced prediction speeds while sustaining high accuracy levels. The progress within deep learning for protein structure prediction has been remarkable, with each successive model surpassing its predecessor. Nonetheless, substantial challenges remain, particularly in achieving accurate predictions for complex proteins.

The protein structure prediction techniques previously examined predominantly rely on multiple sequence alignments. Nevertheless, these methodologies encounter several challenges, including their dependence on homologous sequences, substantial computational demands, and difficulties in addressing complex structural configurations. To tackle these challenges, Wenzhi Mao et al. [54] created a rapid protein structure prediction pipeline by combining the AmoebaContact residue contact predictor with the GDFold folding algorithm. GDFold performs exceptionally well on the PSICOV150 dataset and, although slightly less accurate than RaptorX-Contact on the CASP dataset, it is notably faster. Ratul Chowdhury et al. [55] developed RGN2, a differentiable recurrent geometric network utilizing AminoBERT, for predicting protein structures from sequences. RGN2 surpassed AlphaFold2 and RoseTTAFold in GDT_TS and dRMSD metrics for orphan proteins and was also more computationally efficient and faster. Moreover, Ruidong Wu et al. [56] introduced OmegaFold, a method that accurately predicts high-resolution protein structures from a single sequence. It excels with orphan proteins and antibodies, matching MSA-based methods in CASP and CAMEO evaluations, and surpassing AlphaFold2 and RoseTTAFold with single-sequence inputs. Finally, Zeming Lin et al. [57] developed ESMFold, a large language model that predicts atomic-level protein structures from primary sequences. It operates six times faster than AlphaFold2 on an NVIDIA V100 GPU, facilitating extensive protein structure analysis in metagenomes. These advancements reflect a major shift in protein structure prediction. Methods like GDFold, RGN2, OmegaFold, and ESMFold demonstrate the field's dynamism, each providing unique advantages such as faster processing, enhanced accuracy for specific proteins, or improved computational efficiency. Combining different methods or creating hybrid models could further enhance accuracy and efficiency. To tackle the challenges posed by inadequate sample sizes and limited diversity in existing open-source datasets, Sirui Liu et al. [58] developed the inaugural million-scale protein structure prediction dataset, known as PSP. This dataset secured the first position in the CAMEO competition, thereby offering robust validation of its efficacy. Proteins, being fundamental constituents of life, rely on their three-dimensional conformations for functional and mechanistic analyses. Consequently, sophisticated protein structure prediction methodologies should be employed in drug design and biopharmaceuticals to illustrate their broader potential applications.

### 3.3 Protein Function Prediction

Protein function prediction represents a critical and intricate area of research. The incorporation of deep learning methodologies has introduced novel insights into this domain, substantially improving the precision of predictive outcomes. Contemporary predictive strategies predominantly encompass sequence-based approaches, structural analysis, protein-protein interaction networks, and the integration of multi-source information, as demonstrated in Table II.

***1) Sequence-based protein function prediction:*** Through the analysis of protein amino acid sequences, researchers can infer potential biological functions. This is accomplished by employing methodologies such as homology alignment, functional domain identification, and deep learning techniques. PfmulDL [59] is a protein function annotation method using recurrent and multi-kernel convolutional neural networks with transfer learning to enhance Gene Ontology predictions, especially for "rare class" proteins. Another example is PANDA2 [60], which integrates graph neural networks with evolutionary language models to enhance predictions of protein functions in cellular components and biological processes. Additionally, GOProFormer [61] is a cutting-edge multimodal Transformer that combines protein sequences with the Gene Ontology (GO) hierarchy to enhance protein function prediction accuracy, outperforming previous methods and using new dataset protocols for better model evaluation. Similarly, TEMPROT and TEMPROT+ [62] are Transformer-based frameworks for protein function annotation that improve performance by incorporating BLASTp, particularly for long sequences and rare term predictions. Additionally, the SPROF-GO [63] method uses pre-trained language models for efficient sequence embedding, enhanced by self-attention pooling, hierarchical le-arning, and label diffusion. It outperforms existing methods in accuracy, robustness, and generalization, especially for non-homologous proteins and new species. The HiFun model by Jun Wu et al. [64] translates protein sequences into a "protein language" to predict functions for non-homologous proteins. It surpasses existing methods on the CAFA3 benchmark, particularly for low-homology proteins, and has successfully annotated many previously uncharacterized proteins in the UHGP database, underscoring its practical importance. These advancements suggest a growing trend towards the adoption of sophisticated machine learning techniques for protein function prediction. This shift not only enhances predictive accuracy but also broadens the scope of discovery in the fields of genomics and proteomics.

**TABLE I:** Comparison of Deep Learning Models for Protein Structure Prediction

| Model Name | Network Architecture | Training Dataset | Testing Dataset | Evaluation Indicators | References |
|---|---|---|---|---|---|
| AlphaFold | CNN | PDB、CATH | CASP13 | TM-score = 0.7 | [45] |
| trRosetta | Deep Residual Convolutional Networks | PDB | CASP13 | TM-score = 0.625 | [46] |
| | | | CAMEO | TM-score = 0.621 | |
| ProFOLD | CopulaNet | PDB、CATH | CASP13-FM | TM-score = 0.662 | [47] |
| | | | CASP13-TBM | TM-score = 0.784 | |
| MSA Transformer | Attention mechanism | MSA | CASP13、CAMEO、CB513 | Accuracy = 0.729 | [48] |
| AlphaFold2 | Evoformer | PDB、MSA | CASP14 | TM-score = 0.87 | [13] |
| | | | CAMEO | TM-score = 0.88 | |
| RoseTTAFold | Three-track neural network | PDB、MSA | CASP14 | TM-score = 0.85 | [49] |
| | | | CAMEO | TM-score = 0.82 | |
| DMPfold2 | GRU | CATH | CASP13 | TM-score = 0.590 | [50] |
| ColabFold | MMseqs2 | BFD、MGnify | CASP14 | TM-score = 0.887 | [51] |
| | | | | pLDDT = 88.78 | |
| OpenFold | AlphaFold2 | CATH、OpenProteinSet | CASP15 | GDT-TS = 73.8 | [52] |
| | | | CAMEO | IDDT-Cα = 0.911 | |
| HelixFold | Fused Gated Self-Attention | PDB | CASP14 | TM-score = 0.877 | [53] |
| | | | CAMEO | TM-score = 0.888 | |
| GDFold | AmoebaNet | CATH | PSICOV150、CASP11-13 | TM-score = 0.789 | [54] |
| | | | | r.m.s.d. = 3.66 Å | |
| AminoBERT | RGN2 | UniParc、ProteinNet12、ASTRAL SCOPe | UniRef30、PDB70、MGnify | GDT_TS = 86.5 | [55] |
| OmegaFold | Geoformer | UniRef50 、PDB、SCOP | CASP | TM-score = 0.79 | [56] |
| | | | CAMEO | LDDT = 0.82 | |
| ESMFold | ESM-2 | UniRef | CASP14 | TM-score = 0.68 | [57] |
| | | | CAMEO | TM-score = 0.83 | |
| MEGA-Fold | PSP | PDB、 UniRef50 | CASP14、CAMEO | TM-score = 0.8 | [58] |

***2) Structure-based prediction of protein function:*** Understanding protein structures is crucial for determining their functions. Techniques like X-ray crystallography, NMR, and cryo-electron microscopy, along with computational methods like molecular docking, sequence alignment, and deep learning, enhance the accuracy of functional predictions. For example, DeepFRI [65] is an advanced tool for predicting protein functions, combining a pre-trained LSTM-LM with graph convolutional networks to accurately capture protein structures. It allows high-resolution functional localization via grad-CAM and outperforms on benchmark datasets. Conversely, GAT-GO approach, developed by Boqiao Lai et al. [66], integrates residue contact maps with protein sequence embeddings, thereby enhancing the functional annotation of proteins with low sequence identity and facilitating large-scale annotation efforts. Similarly, EnsembleFam [67] utilizes sequence homology to enhance functional predictions for proteins with low similarity, demonstrating robustness in identifying functions within unknown protein families. Meanwhile, the TransFun [68] model employs pre-trained protein language models and 3D graph neural networks, surpassing current methods on the CAFA3 dataset, with potential for further improvement through sequence similarity integration. Peishun Jiao et al. [69] presented Struct2GO, a graph-based deep learning model demonstrating robust predictive capabilities within the MFO branch, achieving an Fmax of 0.701, AUC of 0.969, and AUPR of 0.796. The HEAL [70] model uses hierarchical graph transformers and contrastive learning to surpass current methods on the PDB test set and AlphaFold2 structures, while improving interpretability through gradient-weighted class activation mapping to pinpoint key functional residues. These methods highlight the evolving approach to predicting protein functions, each with distinct advantages and drawbacks. Some use advanced neural networks for complex features, others combine various data sources or apply ensemble techniques for better accuracy. A major challenge is managing the diverse and complex protein structures and functions.

***3) Protein Function Prediction Based on PPI Networks:*** The structure of PPI networks and traits of central nodes provide key insights into biological processes. Analyzing these networks with deep learning improves the identification and prediction of protein interactions, deepening our understanding of protein functions. The method known as NetQuilt, developed by Barot et al. [71], constructs multi-species protein interaction networks utilizing homologous data to improve the prediction of protein functions. This approach has demonstrated remarkable efficacy in datasets pertaining to humans and mice. Additionally, NetGO 2.0 [72] enhances large-scale protein function prediction by integrating textual annotations with deep sequence data, outperforming its predecessor in predicting biological processes and cellular components, as shown in the CAFA4 challenge. The S2F [73] method enhances protein function prediction in newly sequenced organisms by transferring functional data and employing a label propagation algorithm. It combines homology and protein features, considers overlapping communities in functional networks, and surpasses existing methods, particularly for organisms with experimental data. Moreover, the GLIDER [74] algorithm effectively performs across different PPI networks, emphasizing the need for suitable local similarity measures and optimal k-values for better function prediction. Comparative analyses show variations in network prediction performance. Sai Hu et al. [75] presented RWRT, a tensor-based double random walk model that improves traditional techniques by using functional similarity tensors with multi-omics data. This method identifies more functionally similar partners, minimizes false negatives, and achieves higher accuracy than existing methods on the DIP and BioGRID datasets. Furthermore, the MELISSA [76] method improves GO label prediction by integrating functional labels into embeddings. In large-scale human and yeast multi-network tests, MELISSA outperformed the original Mashup and deepNF methods in predicting protein functions. Overall, these studies have improved protein function prediction, but each method has its pros and cons. Combining data sources can introduce noise and complexity, and more research is needed to generalize these methods across various organisms and systems. Additionally, enhancing the interpretability of deep learning models remains a priority.

***4) Multi-Source Information Fusion for Protein Function Prediction:*** Recently, the field of protein function prediction has evolved from a dependence on single-source data to a more advanced methodology that incorporates information from multiple sources. This paradigm shift encompasses a diverse array of elements, including PPIs, protein domains, amino acid sequences, protein structures, and genomic data. Significant contributions to this field are exemplified by the PFP-GO method developed by Kaustav Sengupta et al. [77], this method combines protein sequences, domain data, and PPI networks to prioritize GO terms, enhancing accuracy and precision metrics while filtering non-essential proteins based on physicochemical properties. Another notable advancement is the ProTranslator [78] framework, which facilitates the prediction of protein functions through the analysis of textual descriptions. This framework efficiently annotates new functions and features in zero-shot and few-shot scenarios, linking genes to biological pathways. Additionally, the DeepGOZero [79] model demonstrates that integrating ontology embedding with neural networks improves protein function prediction, especially for minimally annotated proteins. By leveraging GO axioms to improve embeddings, it attains a 0.903 AUC on the Molecular Function Ontology, outperforming existing methods and supporting zero-shot predictions, demonstrating its practical value. Moreover, Tong Pan et al. [80] effectively utilize attention mechanisms in deep learning for functional annotation with the PfresGO model. By combining protein sequences with gene ontology structures and removing multiple sequence alignments, PfresGO achieves high specificity for GO terms despite low-sequence identity. Similarly, HnetGO [81] combines protein sequences and interaction data to link proteins with GO terms, using pre-trained models for feature extraction. It performs well in cellular component and molecular function categories, highlighting the value of integrating sequence and interaction data for accurate protein function predictions. Lastly, the DeepGATGO [82] model employs a hierarchical pre-training graph attention mechanism, excelling in protein function prediction from input sequences. Its impressive results on CAFA3 and TALE benchmarks demonstrate its scalability and effectiveness. In summary, these advances mark a paradigm shift in protein function prediction, emphasizing the integration of diverse data and innovative computational methods. As research progresses, these multifaceted approaches improve accuracy and deepen insights into protein functionality, paving the way for breakthroughs in understanding biological processes.

### 3.4 Prediction of Protein-Protein Interactions

With the progression of biomedical big data, PPIs have assumed a pivotal role in biological research. Historically, these interactions have been predicted through sequence alignment, structural prediction, and experimental techniques such as mass spectrometry. However, these conventional methods are frequently characterized by their time-intensive nature, high costs, limited generalizability, and susceptibility to high false positive rates. Recent advancements in deep learning have introduced novel methodologies for the prediction of PPIs, as illustrated in Table III.

**TABLE II:** Comparison Table of Deep Learning Models for Protein Function Prediction

| Model Name | Classify | | | | Method | Evaluation Indicators | | | | | | | | Ref |
|---|---|---|---|---|---|---|---|---|---|---|---|---|---|---|
| | Sequence | Structure | PPI | Other | | $F_{max}$ | $S_{min}$ | $F_1$-score | AUC | AUPR | AUROC | AUPRC | Other | |
| PfmulDL | √ | | | | CNN、RNN | √ | | | √ | | | √ | | [59] |
| PANDA2 | √ | | | | GNN、ESM | √ | √ | | | √ | | | | [60] |
| GOProFormer | √ | | | | Transformer | √ | √ | | | | | √ | | [61] |
| TEMPROT | √ | | | | Transformer | √ | √ | | | | | √ | IAuPR | [62] |
| SPROF-GO | √ | | | | Attention | √ | | | | √ | | | | [63] |
| HiFun | √ | | | | CNN<br>Attention<br>BiLSTM | √ | √ | | | √ | | | | [64] |
| DeepFRI | | √ | | | GCN、LSTM | √ | | | | √ | | | | [65] |
| GAT-GO | | √ | | | GAT<br>CNN<br>GNN<br>Attention | √ | | | | | | √ | | [66] |
| EnsembleFam | | √ | | | SVM | | | | √ | | | | ROC | [67] |
| TransFun | | √ | | | EGNN | √ | | | | √ | | | | [68] |
| Struct2GO | | √ | | | GNN<br>Attention | √ | | | √ | √ | | | | [69] |
| HEAL | | √ | | | MPNN<br>HGT<br>Attention | √ | √ | | | √ | | | | [70] |
| NetQuilt | | | √ | | MLP<br>Maxout | | | √ | | | | | MAUPR | [71] |
| NetGO 2.0 | | | √ | | LR-Text<br>Seq-RNN | √ | | | | √ | | | | [72] |
| S2F | | | √ | | InterPro<br>HMMER | √ | √ | | | | | | AUC-ROC<br>AUC-PR | [73] |
| GLIDER | | | √ | | KNN | | | √ | | | | | ACC | [74] |
| RWRT | | | √ | | Double<br>Random<br>Walk<br>Algorithm | | | √ | | | √ | | | [75] |
| MELISSA | | | √ | | ML、CL<br>SVD 、KNN | | | √ | | | | √ | ACC | [76] |
| PFP-GO | | | | √ | Sequence<br>Comparison<br>Algorithms<br>n-Star<br>Consensus<br>Method | | | √ | | | | | Precision<br>Recall | [77] |
| ProTranslator | | | | √ | Transformer | | | | | | √ | | BLEU | [78] |
| DeepGOZero | | | | √ | MLP | √ | √ | | √ | √ | | | | [79] |
| PfresGO | | | | √ | Attention | √ | √ | | | | √ | √ | | [80] |
| HnetGO | | | | √ | Attention | √ | | | √ | √ | | | | [81] |
| DeepGATGO | | | | √ | GAT | √ | | | | | | √ | | [82] |

***1) Based on Deep Neural Networks:*** The development of neural network models enhances the efficient acquisition of structural features and functional information of proteins from extensive biological datasets. These models possess the ability to identify potential interacting protein pairs and elucidate intricate molecular mechanisms. Satyajit Mahapatra et al. [83] created a hybrid model using deep neural networks and extreme gradient boosting to enhance PPI prediction accuracy, excelling in both intra- and inter-species predictions. Additionally, the DWPPI [84] model integrates multi-source data with deep neural networks to precisely predict plant PPIs, achieving high accuracy and AUC in three plant datasets, providing valuable tools for plant molecular biology research. CT-DNN [85] is a PPI prediction method that uses joint trimer encoding and deep neural networks. It effectively processes large protein sequences, automatically extracts features, and enhances prediction accuracy and generalization. Collectively, these models have markedly advanced the prediction of PPIs. The incorporation of deep neural networks, in conjunction with other methodologies, has substantially enriched our comprehension of protein interactions. Nonetheless, despite these advancements, certain areas warrant further investigation. Although the integration of multi-source data proves effective, challenges related to data quality and consistency may arise.

***2) Based on Convolutional Neural Networks:*** CNNs exhibit substantial effectiveness in pattern recognition, facilitating the extraction of feature information from protein sequences and structures. Through the utilization of multi-layered convolutional and pooling processes, these models adeptly identify both local and global features of diverse proteins, thereby improving the accuracy of interaction predictions. Firstly, Xiaotian Hu et al. [86] introduced DeepTrio, a PPI prediction model employing parallel convolutional neural networks. It uses single-protein training with masking to achieve accurate predictions and highlight protein residue importance across various datasets. Secondly, Hongli Gao et al. [87] developed EResCNN, an ensemble residual CNN for PPI prediction, integrating diverse feature extraction methods. It demonstrates high accuracy on datasets from S. cerevisiae, H. pylori, and Human-Yersinia pestis, and performs well in cross-species predictions and PPI network analyses. Finally, Jun Hu et al. [88] created D-PPIsite, a deep learning model for predicting PPI sites, combining convolutional, squeeze-and-excitation, and fully connected layers with four sequence-driven features. It outperforms existing methods on five independent datasets. In summary, CNN-based models have markedly enhanced our comprehension and prediction of protein-protein interactions. The distinctive architectures and methodologies employed by each model contribute unique strengths to the field. As the intricacies of these interaction networks continue to be elucidated, it will be imperative to address the challenges of scalability and interpretability associated with these models.

***3) Based on Recurrent Neural Networks and their variant, Long Short-Term Memory:*** RNNs play a pivotal role in the analysis of protein data due to their capacity to handle sequential information, thereby enhancing our comprehension of protein behavior and interactions. The LSTM variant, which incorporates gating mechanisms to address the vanishing gradient problem, represents a significant advancement in this field. This resulted in models like LSTMPHV [89], which combines LSTM with word2vec embeddings to effectively learn from imbalanced datasets using only amino acid sequences. Its high-precision predictions without biochemical data demonstrate the approach's effectiveness in extracting valuable information from raw sequences. Another notable model, RAPPPID [90] uses a modified AWD-LSTM and various regularization methods for PPI prediction, demonstrating strong performance on the C3 dataset. Its ability to function independently of specific proteins during training and testing underscores its versatility and adaptability for diverse biological applications. Lastly, SENSDeep [91], a sequence-based method for predicting PPI sites, effectively combines deep learning models with innovative features to improve predictive performance. While it may not consistently outperform structure-based methods across all datasets, its efficiency in providing reliable results makes it ideal for quick and dependable predictions. In general, RNN-based models have greatly improved protein research by efficiently processing sequential data and identifying key features, enhancing our understanding of protein interactions and PPI sites. Despite their success with amino acid sequences, incorporating data such as protein structures or post-translational modifications could provide a more complete view.

***4) Based on Graph Neural Networks:*** GNNs represent proteins as nodes and their interactions as edges, allowing them to capture complex interaction features via information propagation. This approach effectively combines structural, functional, and sequential protein data, efficiently handling large biological datasets to enhance predictive accuracy and efficiency. Many research groups have made significant progress in this area. Manon Reau et al. [92] created DeepRank-GNN, a graph neural network framework designed to learn protein-protein interaction patterns. It offers a customizable interface for feature selection, target values, and GNN architectures. The model performed exceptionally well in BM5 and CAPRI benchmarks for scoring docking interactions and distinguishing biological from crystal interfaces. Albu et al. [93] introduced the MM-StackEns method for predicting PPIs, integrating sequence and graph data using Wasserstein distance for feature fusion, significantly enhancing prediction accuracy and generalization with novel protein pairs. Furthermore, Yuting Zhou et al. [94] developed AGAT-PPIS, a PPI site prediction model utilizing an enhanced graph attention network. It incorporates two node features, two edge features, and merges initial residuals with identity mapping, demonstrating strong robustness and generalization across all independent test sets. GNN models have significantly improved our ability to understand and predict protein-protein interactions, opening new avenues for exploring complex protein relationships. Despite their effectiveness in integrating diverse protein data, optimizing graph construction could enhance results. Further insights into information propagation within graphs are essential for advancing and validating these models.

***5) Based on Attention and Transformer:*** The self-attention mechanism is utilized to capture positional relationships within sequences. By leveraging parallel processing and hierarchical structures, the Transformer architecture efficiently handles large-scale protein sequence data, resulting in the generation of high-quality feature representations. For instance, the HANPPIS [95] framework employs a hierarchical attention mechanism with bidirectional gated recurrent units to predict amino acid-level protein interaction sites, offering superior accuracy and interpretability. Similarly, the SDNN-PPI [96] method combines self-attention with deep neural networks, achieving high accuracy across various datasets and PPI network predictions. Furthermore, EnsemPPIS [97] is an ensemble framework that uses transformers and gated convolutional networks for predicting PPI sites. It excels in capturing global and local patterns and residue interactions, demonstrating strong performance across tasks. Its interpretability analysis reveals its ability to learn residue interactions from protein sequences, enhancing predictive accuracy. Generally, self-attention methods have greatly improved protein research by enhancing our ability to predict interaction sites and understand amino acid relationships. Despite their accuracy, these models are computationally intensive with large protein datasets. Incorporating biological knowledge like post-translational modifications and protein folding could further boost predictions.

***6) Based on Autoencoders:*** Given the expanding scale of protein datasets, autoencoders employ unsupervised learning methodologies to distill critical features. This process facilitates the efficient compression of protein sequence or structural data into low-dimensional representations, which encapsulate potential interaction patterns. The AutoPPI method, as developed by Gabriela Czibula et al. [98], exhibits remarkable efficacy in predicting protein-protein interactions by employing a deep autoencoder, thereby attaining high accuracy and AUC metrics across diverse datasets. In parallel, The DHL-PPI model by Yue Jiang et al. [99] uses deep learning to transform protein sequences into binary hash codes, enabling efficient protein-protein interaction prediction via Hamming distance. It achieves high precision, recall, and F1-score across datasets from four species, while reducing computational complexity. Meanwhile, the ProtInteract framework by Farzan Soleymani et al. [100] uses autoencoders for protein simplification and deep CNNs for PPI prediction. Experiments reveal that TCNs surpass LSTMs in both accuracy and speed, making them preferable for large-scale analyses due to reduced computational requirements. Autoencoder-based methods have significantly advanced protein research by enabling efficient analysis of large datasets and improving protein-protein interaction predictions. However, challenges remain, such as potential loss of detailed information when compressing data into low-dimensional representations. Future studies should aim to balance data compression with the preservation of essential details.

## IV. CHALLENGES, LIMITATIONS, AND RISKS

Despite significant advancements in the application of deep learning technologies in the field of proteomics, numerous challenges remain.

***1) Data:*** Deep learning models require large, high-quality datasets for effective training, but proteomics research often struggles with missing values and noise due to experimental methods and biological sample variability, hindering model performance. Furthermore, the low expression levels of key proteins within samples contribute to data imbalance issues [101]. These factors underscore the need for data cleaning, preprocessing, augmentation, and regularization to accurately reflect biological phenomena.

***2) Calculation:*** In the domain of proteomics, the deployment of deep learning models generally necessitates significant computational resources, especially when processing large-scale and high-dimensional biological datasets [102]. Traditional hardware often struggles to efficiently train models, limiting their use. Integrating cloud and high-performance computing, along with improved model compression and acceleration algorithms, is expected to boost training efficiency and reduce time.

***3) Interpretability:*** The intricate hidden layers and "black box" characteristics of deep learning

**TABLE III:** Comparison Table of Deep Learning Models for Protein-Protein Interaction Prediction

| Classify | Model Name | Dataset | Evaluation Indicators (%) | | | | | | | | | | | | Ref |
|---|---|---|---|---|---|---|---|---|---|---|---|---|---|---|---|
| | | | ACC | Sens | Spec | Prec | MCC | AUC | $F_1$-score | AUPR | AUROC | Recall | AUPRC | Other | |
| DNN | DNN-XGB | Saccharomyces cerevisiae | 98.35 | 97.78 | 98.93 | 98.91 | 96.72 | 99.70 | / | / | / | / | / | / | [83] |
| | | Helicobacter pylori | 96.19 | 96.71 | 95.68 | 95.76 | 92.41 | 98.60 | | | | | | | |
| | | Human-Bacillus Anthracis | 98.50 | 98.67 | 98.31 | 98.33 | 97.00 | 99.87 | | | | | | | |
| | | Human-Yersinia pestis | 97.25 | 97.68 | 97.68 | 96.86 | 94.51 | 99.29 | | | | | | | |
| | DWPPI | Arabidopsis thaliana | 89.47 | 91.47 | 87.48 | 87.97 | 79.02 | 95.48 | / | / | / | / | / | / | [84] |
| | | Zea mays | 95.00 | 96.30 | 93.69 | 93.85 | 90.02 | 98.67 | | | | | | | |
| | | Oryza sativa | 85.63 | 86.38 | 84.89 | 85.11 | 71.28 | 92.13 | | | | | | | |
| | CT-DNN | HPRD | / | / | / | / | / | 98.3 | / | 98.2 | / | / | / | / | [85] |
| | | Swiss-Prot | | | | | | | | | | | | | |
| CNN | DeepTrio | BioGRID S.cerevisiae | 97.55 | 96.12 | 98.98 | 98.95 | 95.15 | / | 97.52 | / | / | / | / | / | [86] |
| | | BioGRID H.sapiens | 98.12 | 97.23 | 99.01 | 99.00 | 96.26 | | 98.11 | | | | | | |
| | | DeepFE-PPI S.cerevisiae | 92.57 | 88.53 | 96.62 | 96.33 | 92.26 | | 85.43 | | | | | | |
| | | PIPR S.cerevisiae | 94.78 | 92.20 | 97.33 | 97.18 | 94.63 | | 89.67 | | | | | | |
| | EResCNN | S. cerevisiae | 95.34 | / | / | 98.32 | 90.86 | / | / | / | / | 92.26 | / | / | [87] |
| | | H. pylori | 87.89 | | | 87.84 | 75.81 | | | | | 87.96 | | | |
| | | Human-Y. pestis | 98.61 | | | 97.56 | 97.23 | | | | | 98.65 | | | |
| | D-PPIsite | Dset_186 | 80.9 | 37.3 | 88.7 | 37.3 | 26.0 | 73.2 | 37.3 | 35.7 | / | / | / | / | [88] |
| | | Dset_72 | 85.1 | 29.9 | 91.7 | 29.9 | 21.6 | 74.0 | 29.9 | 25.4 | | | | | |
| | | Dset_164 | 77.8 | 38.6 | 86.4 | 38.6 | 25.0 | 71.0 | 38.6 | 36.4 | | | | | |

| | | | | | | | | | | | | | | | | |
|---|---|---|---|---|---|---|---|---|---|---|---|---|---|---|---|---|
| | | Dset_448 | | 85.9 | 48.1 | 91.9 | 48.0 | 39.9 | 82.4 | 48.0 | 47.9 | | | | | |
| | | Dset_355 | | 87.1 | 46.0 | 92.7 | 46.0 | 38.7 | 82.2 | 46.0 | 44.8 | | | | | |
| RNN & LSTM | LSTM-PHV | Training Dataset | | 98.4 | / | / | / | / | 97.6 | / | / | / | / | / | / | [89] |
| | | Independent Dataset | | 98.5 | | | | | 97.3 | | | | | | | |
| | | SARS-CoV-2 PPIs | | / | | | | | 95.6 | | | | | | | |
| | | Non-viral pathogens PPIs | | | | | | | 92.2 | | | | | | | |
| | RAPPPID | STRING | C1 | / | / | / | / | / | 97.8 | / | 97.4 | 97.8 | / | / | / | [90] |
| | | | C2 | | | | | | 85.9 | | 86.8 | 85.9 | | | | |
| | | Negatome | C3 | | | | | | 80.3 | | 81.0 | 80.3 | | | | |
| | SENSDeep | Dset_186 | | 80.7 | 38.8 | 88.3 | 37.6 | 26.8 | / | 38.2 | / | 72.5 | / | 35.0 | / | [91] |
| | | Dset_72 | | 80.8 | 40.4 | 86.1 | 27.4 | 22.6 | | 32.7 | | 71.5 | | 26.5 | | |
| | | Dset_164 | | 78.9 | 30.9 | 89.2 | 38.0 | 21.8 | | 34.1 | | 68.6 | | 33.9 | | |
| | | Dset_448 | | 83.2 | 29.8 | 91.7 | 36.6 | 23.5 | | 32.8 | | 68.1 | | 31.0 | | |
| | | Dset_355 | | 84.8 | 30.7 | 92.2 | 34.9 | 24.2 | | 32.6 | | 69.0 | | 29.7 | | |
| | DeepRank-GNN | BM5 | | / | / | / | / | / | 85 | / | / | / | / | / | Success rates = 93.3 | [92] |
| | | CAPRI | | | | | | | 71 | | | | | | Success rates = 76.9 | |
| | | DC | | 82 | 83 | 81 | 82 | | / | | | | | | / | |
| | | Yeast | C1 | 81 | | | 78 | | | | 92 | 91 | 89 | | | |
| | | | C2 | 69 | | | 69 | | | | 77 | 76 | 71 | | | |
| | | | C3 | 64 | | | 71 | | | | 71 | 69 | 47 | | | |
| GNN | MM-StackEns | Human | C1 | 78 | / | / | 72 | / | / | / | 88 | 88 | 89 | / | / | [93] |
| | | | C2 | 64 | | | 61 | | | | 70 | 71 | 75 | | | |
| | | | C3 | 62 | | | 66 | | | | 68 | 68 | 50 | | | |
| | | Human-2021 | 50% Random | 85 | | | 80 | | | | 92 | 92 | 92 | | | |
| | | | 10% Random | 89 | | | 46 | | | | 70.5 | 91 | 80 | | | |
| | | | 0.3% Random | 90 | | | 30 | | | | 13.6 | 91 | 81 | | | |
| | | Yeast-2017 | | 94.0 | | | 94.5 | | | | 98.3 | 98.1 | 93.3 | | | |
| | | Multi-species | <40% | 57.5 | | | 57.2 | | | | 61.3 | 59.0 | 85.3 | | | |
| | | | <25% | 58.5 | | | 58.5 | | | | 62.8 | 59.6 | 85.0 | | | |
| | | | <10% | 59.2 | | | 59.7 | | | | 65.1 | 60.5 | 84.6 | | | |
| | | | <1% | 59.3 | | | 60.2 | | | | 65.6 | 60.2 | 85.0 | | | |
| | AGAT-PPIS | Test_60 | | 85.6 | / | / | 53.9 | 48.4 | / | 56.9 | / | 86.7 | 60.3 | 57.4 | / | [94] |
| | | Test_315-28 | | / | | | / | 48.1 | | / | | / | / | 57.2 | | |
| | | Btest_31-6 | | | | | | 48.5 | | | | | | 58.3 | | |
| | | UBtest_31-6 | | | | | | 32.7 | | | | | | 36.5 | | |
| Attention & Transformer | HANPPIS | Dset_186 | | 63.1 | / | / | 29.1 | / | / | 39.3 | / | / | 60.5 | / | / | [95] |
| | | Dset_72 | | | | | | | | | | | | | | |
| | | Dset_164 | | | | | | | | | | | | | | |
| | SDNN-PPI | S.cerevisiae | | 95.48 | 93.80 | 97.23 | 97.13 | 91.02 | 98.63 | / | / | / | / | / | / | [96] |
| | | Human | | 98.94 | 98.77 | 99.10 | 99.02 | 97.57 | 99.60 | | | | | | | |
| | | Human-B.Anthracis | | 93.15 | 96.61 | 89.69 | 90.44 | 86.57 | 98.23 | | | | | | | |
| | | Human-Y.pestis | | 88.33 | 93.92 | 82.74 | 84.63 | 77.26 | 95.74 | | | | | | | |
| | EnsemPPIS | DeepPPISP task | | 73.2 | / | / | 37.5 | 27.7 | / | 44.0 | / | 71.9 | 53.2 | 40.5 | / | [97] |
| | | DELPHI task | | 82.1 | | | / | 29.1 | | 38.5 | | 77.0 | / | 35.4 | | |
| AE | AutoPPI | HPRD | Joint–Joint | 97.7 | / | 98.6 | 98.6 | / | 97.7 | 97.7 | / | / | 96.8 | / | / | [98] |
| | | | Siamese–Joint | 97.9 | | 97.3 | 97.3 | | 97.9 | 97.9 | | | 98.5 | | | |
| | | | Siamese–Siamese | 96 | | 99.2 | 99.2 | | 96.0 | 95.9 | | | 92.8 | | | |
| | | Multi-species | Joint–Joint | 97 | | 99.5 | 99.5 | | 97 | 96.9 | | | 94.4 | | | |
| | | | Siamese–Joint | 96.9 | | 96.4 | 96.5 | | 97 | 97 | | | 97.4 | | | |
| | | | Siamese–Siamese | 98.2 | | 100 | 100 | | 98.2 | 98.2 | | | 96.4 | | | |
| | | Multi-species < 0.25 | Joint–Joint | 97.3 | | 99.5 | 99.5 | | 97.5 | 97.5 | | | 95.6 | | | |
| | | | Siamese–Joint | 97.6 | | 96.8 | 97.4 | | 97.5 | 97.8 | | | 98.3 | | | |
| | | | Siamese–Siamese | 98.3 | | 100 | 100 | | 98.5 | 98.4 | | | 96.9 | | | |
| | | Multi-species < 0.01 | Joint–Joint | 97.2 | | 99.1 | 99.3 | | 97.5 | 97.5 | | | 95.8 | | | |
| | | | Siamese–Joint | 97.8 | | 96.6 | 97.5 | | 97.6 | 98.1 | | | 98.7 | | | |
| | | | Siamese– | 98.1 | | 100 | 100 | | 98.3 | 98.3 | | | 96.6 | | | |
| | | | Siamese | | | | | | | | | | | | | |
| | DHL-PPI | C. elegans | | 98.8 | / | 100 | 100 | 97.5 | / | 99.0 | / | / | 98.1 | / | / | [99] |
| | | Drosophila | | 98.8 | | 99.7 | 99.8 | 97.6 | | 99.0 | | | 98.1 | | | |
| | | E. coli | | 97.1 | | 98.2 | 98.7 | 94.0 | | 97.5 | | | 96.2 | | | |
| | | Human | | 97.1 | | 98.0 | 98.4 | 94.1 | | 97.3 | | | 96.3 | | | |
| | ProtInteract | H. sapiens | 2- classes | 95.68 | / | / | 95.50 | / | 96.00 | / | / | / | / | / | / | [100] |
| | | | 3- classes | 92.44 | | | 89.50 | | 93.40 | | | | | | | |
| | | | 5- classes | 91.32 | | | 79.20 | | 90.60 | | | | | | | |
| | | M. musculus | 2- classes | 91.45 | | | 84.50 | | 86.00 | | | | | | | |
| | | | 3- classes | 89.83 | | | 78.70 | | 86.00 | | | | | | | |
| | | | 5- classes | 87.49 | | | 70.00 | | 84.80 | | | | | | | |

models significantly complicate the interpretability of their decision-making processes and outcomes [103]. In medicine and biological research, interpretability is crucial for understanding a model's conclusions, ensuring safe and effective use. Researchers are exploring strategies to enhance interpretability, such as developing explanatory tools, visualization techniques, and knowledge-based methods. These efforts aim to improve deep learning model transparency, fostering better understanding and use.

***4) Transparency:*** The transparency of deep learning models is crucial, especially in biomedical applications, due to the sensitivity of medical data [104] which often prevents open-sourcing. Enhancing model transparency can help clinicians understand decision-making processes and build trust in the results.

***5) Ethics:*** Proteomics research uses data from patient samples, making privacy protection and ethics crucial due to risks of genetic information exposure and discrimination. Future research should focus on creating secure data processing and storage technologies to protect privacy and ensure security, while adhering to ethical guidelines for legal and moral integrity.

## V. FUTURE DIRECTIONS

In the domain of proteomics informatics leveraging deep learning, prospective research trajectories of significant interest encompass multimodal data fusion, self-supervised learning [105], and transfer learning [106]. Additionally, fostering interdisciplinary collaboration and the integration of diverse knowledge bases are crucial areas for future exploration.

Integrating multimodal data is expected to become crucial in proteomics research, enhancing traditional proteomic data from techniques like mass spectrometry and NMR with genomics, transcriptomics, and metabolomics. Deep learning technologies effectively integrate diverse data sources, uncovering key insights into protein expression, function, and interactions. This progress enhances biomarker identification and deepens our understanding of disease mechanisms.

Implementing self-supervised and transfer learning is expected to significantly boost model effectiveness and applicability. Self-supervised learning leverages unlabeled data for feature representation, performing well in data-scarce situations. Meanwhile, transfer learning applies existing knowledge by using model weights from other biological tasks in proteomics, speeding up training and optimization.

Interdisciplinary collaboration and diverse knowledge integration are key to advancing proteomic informatics. By uniting experts in biology, computer science, physics, and statistics, a more comprehensive approach can be developed to understand proteins' complex roles in biological processes. Significant progress in the speed of proteomics research and its future applications can be made by encouraging interdisciplinary networks that support data sharing and collaborative knowledge creation, along with the integration of advanced technologies and theoretical models.

## VI. CONCLUSIONS

This research reviews recent advancements in proteomics informatics using deep learning, highlighting the effectiveness of various algorithms in predicting protein sequences, structures, functions, and interactions. These technologies have significantly enhanced the predictive accuracy of proteomic research and laid a strong foundation for future studies. Despite these advancements, challenges remain. Future research should focus on improving the transparency and interpretability of deep learning models to further our understanding of proteomics informatics.

## ACKNOWLEDGEMENTS

This work is supported by the Natural Science Foundation of Fujian Province (2023J05083, 2022J011396, 2023J011434).